# On the Structure of Equilibrium Strategies in Dynamic Gaussian Signaling Games

Muhammed O. Sayin, Emrah Akyol, Tamer Başar

*Abstract*— This paper analyzes a finite horizon dynamic signaling game motivated by the well-known strategic information transmission problems in economics. The mathematical model involves information transmission between two agents, a sender who observes two Gaussian processes, state and bias, and a receiver who takes an action based on the received message from the sender. The players incur quadratic instantaneous costs as functions of the state, bias and action variables. Our particular focus is on the Stackelberg equilibrium, which corresponds to information disclosure and Bayesian persuasion problems in economics. Prior work solved the static game, and showed that the Stackelberg equilibrium is achieved by pure strategies that are linear functions of the state and the bias variables. The main focus of this work is on the dynamic (multi-stage) setting, where we show that the existence of a pure strategy Stackelberg equilibrium, within the set of linear strategies, depends on the problem parameters. Surprisingly, for most problem parameters, a pure linear strategy does not achieve the Stackelberg equilibrium which implies the existence of a trade-off between exploiting and revealing information, which was also encountered in several other asymmetric information games.

## I. INTRODUCTION

This paper focuses on signaling games where information transmission occurs between a *sender* who observes the state perfectly, and a *receiver* who takes actions based on the received signals from the sender. The agents have misaligned objectives that are functions of the state and the actions. The literature on such games typically focuses on the settings where the sender transmits exogenous information to the receiver [1], [2]. Recent pioneering works on Bayesian persuasion [3] and optimal information disclosure [4] depart from this tradition by instead asking what information the sender would generate if he initially is as uninformed as the receiver and commits to fully revealing the output of a *transmission* strategy that is determined based on the statistics, ex-ante. More formally, while the classical prior work has focused on the Nash equilibrium, recent works analyze the Stackelberg equilibrium of the same problem setting. Applications range from prosecutors gathering evidence for presentation at court (a prosecutor-sender trying the persuade a judge-receiver for a guilty verdict, by carefully selecting which tests-transmission strategies to apply) to specifying the terms of free trials of recently developed products. Due to wide applicability of such mathematical models, there has been a growing interest in such problems in the economics literature, see e.g., [3]–[12], and very recently in control [13], [14]. Reference [13] has studied networked estimation with biased sensors in the context of Stackelberg equilibrium, and in [14], the authors have extended the approach of [1] to noisy and multidimensional settings.

The commitment of the agents yields radically different conclusions in terms of optimal strategies. For example, while it is well understood that optimal strategies achieving Nash equilibria are non-injective, the Stackelberg equilibrium admits linear optimal strategies for settings involving Gaussian variables and quadratic cost functions [15]. Sender's commitment is also essential for the derivation of the fundamental bounds in communication using information theory, which models the problem in terms of ex-ante mappings that are designed based on statistics, see [16] for more details.

Computation of optimal strategies for general dynamic Gaussian-quadratic games with asymmetric information has been an active research area, see e.g., [17] and the references therein. In this paper, we focus on a particular finite horizon dynamic signaling game with two players: a sender and a receiver. The sender observes two primitive random processes (the state and the bias) and sends a signal to the receiver who takes actions based on the received signal. The state and the bias processes evolve as independent (uncontrolled) Gaussian processes, and players incur quadratic instantaneous costs. The static version of this problem was solved in [15], [16] (see also [18]). The main conclusion of the static game is that the Stackelberg equilibrium admits essentially unique pure strategies that are linear functions of the state and the bias variables. It is intuitively expected that a similar conclusion would hold in the dynamic version of the same game. Here, we show that this intuition is correct for only very specific problem parameters (when the strategies are restricted to be linear), while, in general, the equilibrium is achieved by mixed strategies which imply that the sender adds noise to its signal at the equilibrium. This result can be interpreted as follows: In the first stage of the game, the optimal sender strategy is to transmit a linear function of the state and the bias variable. But for most problem parameters (which we characterize precisely in Section IV), this pure strategy reveals too much to the receiver about the sender variables available at the second stage, and is suboptimal for the second stage performance of the sender. Hence, the sender faces a trade-off between the first and second stages of this game in terms of his strategies. This trade-off results in mixed strategies that are equivalent, when constrained to

This research was supported in part by the U.S. Air Force Office of Scientific Research (AFOSR) MURI grant FA9550-10-1-0573, in part by the U.S. Office of Naval Research (ONR) MURI grant N00014-16-1-2710, and in part by NSF under grant CCF 11-11342.

Authors are with the Coordinated Science Laboratory, University of Illinois at Urbana-Champaign, Urbana, IL. 61801; emails:{sayin2, akyol, basar1}@illinois.edu

linear strategies, to adding independent noise to the sender's signal to hide information about the second stage. In passing, we note that similar trade-offs were also observed in more general asymmetric information games where exploiting the available information comes with the cost of revealing it, see e.g., [19].

The paper is organized as follows: In Section II, we provide preliminaries. In Section III, we investigate the existence of a Nash equilibrium. In Section IV, we present our main result on the Stackelberg equilibrium. In Section V, we discuss contributions and future research directions.

## II. PRELIMINARIES

Consider the two discrete-time, stationary, exogenous processes $\{x_k\}$, the state, and $\{\theta_k\}$, the bias, that evolve according to the following first-order autoregressive models

$$x_{k+1} = A_x x_k + w_k,$$
$$\theta_{k+1} = A_\theta \theta_k + v_k, \quad k = 1, 2, \cdots, \quad (1)$$

where $A_x, A_\theta \in \mathbb{R}^{p \times p}$, $x_1 \sim \mathbb{N}(0, \Sigma_x)$, $\theta_1 \sim \mathbb{N}(0, \Sigma_\theta)$, and $x_1 \perp\!\!\!\perp \theta_1$.[1] The additive noise processes $\{w_k\}$ and $\{v_k\}$ are white Gaussian random processes, i.e., $w_k \sim \mathbb{N}(0, \Sigma_w)$ and $v_k \sim \mathbb{N}(0, \Sigma_v)$, which are independent of each other, and of the current state $x_k$ and bias $\theta_k$.

Let there be two decision makers: a sender and a receiver, which take actions according to different objectives. Here, only the sender has access to the state and the bias variables. In particular, at time instant $k$, the sender has access to $x_1^k$ and $\theta_1^k$, where, by a possible abuse of notation, we define the augmented vectors $x_j^k := [x_j' \cdots x_k']'$ and $\theta_j^k := [\theta_j' \cdots \theta_k']'$.

Sender's actions $y_k$, for $k = 1, 2, \cdots$, are given by

$$y_k = \eta_k(x_1^k, \theta_1^k, z_1^k)$$
$$= \sum_{j=1}^{k} B_{k,j} x_j + C_{k,j} \theta_j + D_{k,j} z_j, \quad (2)$$

where $B_{k,j}, C_{k,j}, D_{k,j} \in \mathbb{R}^{r \times p}$ are design parameters, $z_k \sim \mathbb{N}(0, I)$ for $k = 1, 2, \cdots$, are multivariate Gaussian random variables that are independent of each other and of the other parameters, and $z_1^k = [z_1' \cdots z_k']'$. Particularly, the sender chooses the policy[2] $\eta_k(\cdot, \cdot, \cdot)$ from the policy space $\Omega_k$, which is the set of linear functions that map from $\mathbb{R}^{kp} \times \mathbb{R}^{kp} \times \mathbb{R}^{kp}$ to $\mathbb{R}^r$, i.e., $\eta_k(\cdot, \cdot, \cdot) \in \Omega_k$.

Even though the receiver does not have access to the state and bias parameters, he/she has access to the actions of sender, and takes actions $u_k$, are given by

$$u_k = \gamma_k(y_1, \cdots, y_k), \quad (3)$$

for $k = 1, 2, \cdots$, and the policy $\gamma_k(\cdot)$ is chosen from the policy space $\Gamma_k$ which is the set of all Lebesgue measurable functions from $\mathbb{R}^{kr}$ to $\mathbb{R}^p$, i.e., $\gamma_k(\cdot) \in \Gamma_k$.

---
[1] Here, $\mathbb{R}^{p \times p}$ denotes the space of $p \times p$ matrices with real-valued entries, $\mathbb{N}(0, .)$ denotes the multivariate Gaussian distribution with zero mean and designated covariance, and $X \perp\!\!\!\perp Y$ means that the random variables $X$ and $Y$ are independent.

[2] In this paper, we use the terms "policy" and "strategy" interchangeably.

The sender wants to minimize the following finite horizon objective[3]

$$J_T(\eta_1, \cdots, \eta_n; \gamma_1, \cdots, \gamma_n) = \sum_{k=1}^{n} \mathbb{E}\{\|x_k + \theta_k - u_k\|^2\} \quad (4)$$

over $\eta_k(\cdot, \cdot, \cdot) \in \Omega_k$, for $k = 1, \cdots, n$. The objective of the receiver is to minimize a different function,

$$J_R(\eta_1, \cdots, \eta_n; \gamma_1, \cdots, \gamma_n) = \sum_{j=1}^{n} \mathbb{E}\{\|x_k - u_k\|^2\} \quad (5)$$

over $\gamma_k(\cdot) \in \Gamma_k$, for $k = 1, \cdots, n$.

## III. NASH EQUILIBRIUM

We first consider the case in which the agents announce their policies simultaneously. This corresponds to a Nash equilibrium problem [20] such that in the equilibrium, no agent has any incentive to change his/her policy unilaterally. In particular, equilibrium achieving policy pairs $(\eta^*, \gamma^*)$, where $\eta^* := (\eta_1^*, \cdots, \eta_n^*)$ and $\gamma^* := (\gamma_1^*, \cdots, \gamma_n^*)$, minimize $J_T(\eta_1, \cdots, \eta_n; \gamma_1^*, \cdots, \gamma_n^*)$ and $J_R(\eta_1^*, \cdots, \eta_n^*; \gamma_1, \cdots, \gamma_n)$, respectively.

For each $n$-tuple of policies $\eta_k \in \Omega_k$, $k = 1, \cdots, n$, let $R_R(\eta_1, \cdots, \eta_n)$ be the reaction set of the receiver:

$$R_R(\eta_1, \cdots, \eta_n) := \arg \min_{\substack{\gamma_k \in \Gamma_k, \\ k=1,\cdots,n}} J_R(\eta_1, \cdots, \eta_n; \gamma_1, \cdots, \gamma_n). \quad (6)$$

Correspondingly, for each $n$-tuple of policies $\gamma_k \in \Gamma_k$, $k = 1, \cdots, n$, the reaction set of the sender is given by

$$R_T(\gamma_1, \cdots, \gamma_n) := \arg \min_{\substack{\eta_k \in \Omega_k, \\ k=1,\cdots,n}} J_T(\eta_1, \cdots, \eta_n; \gamma_1, \cdots, \gamma_n).$$

This implies that the equilibrium achieving policy pairs $(\eta^*, \gamma^*)$ satisfy

$$\eta^* \in R_T(\gamma^*) \text{ and } \gamma^* \in R_R(\eta^*). \quad (7)$$

However, existence of a Nash equilibrium point is not guaranteed a priori, as we further discuss below.

We note that reference [1] has studied single-stage, i.e., $n = 1$, strategic information transmission of scalar source and bias information under Nash equilibrium. In that problem, the policy space of the transmitter is the set of all Lebesgue measurable functions of the state and the bias information. In that general single-stage context, it was shown that all the equilibrium achieving policies of the transmitter are quantization based, i.e., non-injective, mappings of the state and the bias information. In particular, linear policies for the transmitter do not achieve a Nash equilibrium.

Different from [1], however, here the receiver has no direct access to the bias information $\theta_k$, whereas in [1], the bias was a common knowledge of both agents. In our framework, the following theorem addresses the existence of Nash equilibrium.

**Theorem 1.** *In both single and multi-stage problems with the objectives* (4) *and* (5), *a (pure-strategy) Nash equilibrium* (7) *does not exist.*

---
[3] $\|\cdot\|$ denotes the Euclidean ($L^2$) norm of $(\cdot)$.

*Proof.* We point out that the optimization problem (5) faced by the receiver is a standard recursive estimation problem in which minimizing policies $\gamma_k$, $k = 1, \cdots, n$, are unique for each $\eta_k$. In particular, for a given $k$-tuple $\eta_1, \cdots, \eta_k$, the optimum policy $\gamma_k^*$ is given by

$$\gamma_k^*(y_1, \cdots, y_k; \eta_1, \cdots, \eta_k) = \mathbb{E}\{x_k | y_1 = \eta_1(x_1, \theta_1, z_1), \cdots, y_k = \eta_k(x_1^k, \theta_1^k, z_1^k)\}. \quad (8)$$

Since $(x_1^k, \theta_1^k, z_1^k)$ are jointly Gaussian for each $\eta_k \in \Omega_k$ and have zero-mean, $\gamma_k^*$ is going to be linear in $(y_1, \cdots, y_k)$. Hence, for Nash equilibrium, (8) yields that $\gamma_k^*$, $k = 1, \cdots, n$, are going to be unique and linear in the actions of sender, i.e., $(y_1, \cdots, y_k)$. Assume now that the policy pair $(\eta^*, \gamma^*)$ achieves a Nash equilibrium. Then, by (8), we can write

$$\gamma_k^*(y_1, \cdots, y_k; \eta_1, \cdots, \eta_k) = \sum_{j=1}^{k} \kappa_{k,j}(\eta^*) y_j \quad (9)$$

and for given $\kappa_{k,j}^* = \kappa_{k,j}(\eta^*)$, $k = 1, \cdots, n$, and $j = 1, \cdots, k$, the sender faces the following optimization problem

$$\min_{\substack{\eta_k \in \Omega_k, \\ k=1\cdots,n}} J_T\left(\eta_1, \cdots, \eta_n; \gamma_1 = \kappa_{1,1}^* y_1, \cdots, \gamma_n = \sum_{j=1}^{n} \kappa_{n,j}^* y_j\right),$$

$$= \min_{\substack{\eta_k \in \Omega_k, \\ k=1\cdots,n}} \sum_{k=1}^{n} \mathbb{E}\left\{\left(x_k + \theta_k - \sum_{j=1}^{k} \kappa_{k,j}^* y_j\right)^2\right\}, \quad (10)$$

where

$$y_j = \sum_{i=1}^{j} b_{j,i} x_i + c_{j,i} \theta_i + d_{j,i} z_i \quad (11)$$

and we use $b_{k,j}, c_{k,j}, d_{k,j} \in \mathbb{R}$ instead of $B_{k,j}, C_{k,j}, D_{k,j}$ from (2) since we consider the scalar case. This quadratic objective function is continuous in parameters and bounded from below, i.e., non-negative.

Note that by (8), (9) and (11), we have $\kappa_{kk}^* \neq 0$ if $b_{k,k} \neq 0$. Then, $b_{k,k} = c_{k,k} = (\kappa_{k,k}^*)^{-1}$,

$$b_{k,j} = -\frac{1}{\kappa_{k,k}^*} \sum_{i=1}^{k-1} \kappa_{k,i}^* b_{i,j}, \quad c_{k,j} = -\frac{1}{\kappa_{k,k}^*} \sum_{i=1}^{k-1} \kappa_{k,i}^* c_{i,j},$$

and $d_{k,k} = d_{k,j} = 0$ for $k = 1, \cdots, n$ and $j = 1, \cdots, k-1$ achieve the global minimum in (27), which is zero. Let $\tilde{\eta}^* = (\tilde{\eta}_1^*, \cdots, \tilde{\eta}_n^*)$ be the corresponding $n$-tuple of optimum policies. By (9) and (11), $\tilde{\eta}^*$ implies that $\gamma_1^*$ is given by

$$\gamma_1^*(y_1; \tilde{\eta}_1^*) = \kappa_{1,1}^* \left(\frac{1}{\kappa_{1,1}^*} x_1 + \frac{1}{\kappa_{1,1}^*} \theta_1\right), \quad (12)$$

which yields $\gamma_1^*(y_1; \tilde{\eta}_1^*) = x_1 + \theta_1$. However, by (3) and (5), for given $\tilde{\eta}^*$, $\gamma_1$ minimizes the following sub-problem:

$$\min_{\gamma_1 \in \Gamma_1} \mathbb{E}\left\{\left(x_1 - \gamma_1\left(\frac{1}{\kappa_{1,1}^*} x_1 + \frac{1}{\kappa_{1,1}^*} \theta_1\right)\right)^2\right\}$$

and the corresponding optimum policy is given by

$$\gamma_1^*\left(\frac{1}{\kappa_{1,1}^*} x_1 + \frac{1}{\kappa_{1,1}^*} \theta_1; \tilde{\eta}_1^*\right) = \frac{\sigma_x^2}{\sigma_x^2 + \sigma_\theta^2}(x_1 + \theta_1), \quad (13)$$

which is not equal to (12) for any statistics of the parameters. In particular, the receiver always has an incentive to change his/her policy (12) to (13). This yields a contradiction for both single and multi stage Nash equilibria (7), therefore there are no policy pairs that lead to a pure-strategy Nash equilibrium. $\square$

Since our main interest in this paper is the Stackelberg equilibrium (discussed next), we do not pursue existence of mixed-strategy Nash equilibrium or pure-strategy equilibrium in nonlinear policies.

## IV. STACKELBERG EQUILIBRIUM

We now consider that there is a hierarchy between the agents in the announcement of the policies such that the sender leads the game by announcing his/her policies beforehand and sticks to that announced policy. This corresponds to a Stackelberg equilibrium problem [20] such that the leader, i.e., the sender, chooses his policy based on the corresponding best response of the follower, i.e., the receiver. In our problem, by (6) and (8), for each $\eta_k \in \Omega_k$, the reaction set of the receiver is a singleton since the corresponding best response policies of the receiver are unique and given by (8). To this end, the sender seeks optimal policies $\eta^*$ which minimize $J_T(\eta_1, \cdots, \eta_n; \gamma_1, \cdots, \gamma_n)$ over $\eta_k \in \Omega_k$ and $\gamma_k \in R_R(\eta_1, \cdots, \eta_n)$ for $k = 1, \cdots, n$. In particular, the optimization problem faced by the sender is given by

$$\min_{\substack{\eta_k \in \Omega_k, \\ k=1,\cdots,n}} J_T(\eta_1, ..., \eta_n; \gamma_1^*(y_1; \eta_1), ..., \gamma_n^*(y_1, ..., y_n; \eta_1, ..., \eta_n)) \quad (14)$$

and for $k = 1, \cdots, n$, $\gamma_k^*$ is given by (8).

**Proposition 1.** *The Stackelberg equilibrium problem* (14) *attains a global minimum.*

*Proof.* The optimization objective in (14) is continuous in the design parameters, i.e., $b_{k,j}, c_{k,j}, d_{k,j} \in \mathbb{R}$, $j = 1, \cdots, k$, that define $\eta_k$ in (2), and bounded from below since the quadratic measure (4) is non-negative. Then, consider that the design parameters $b_{k,j}, c_{k,j}, d_{k,j} \in \bar{\mathbb{R}}$, $k = 1, \cdots, n$, and $j = 1, \cdots, k$, where $\bar{\mathbb{R}} = \mathbb{R} \cup \{-\infty, +\infty\}$ is the extended real line. For this problem, since the design parameters can take values over the compact set $\bar{\mathbb{R}}$ and the objective function is continuous in the design parameters, (14) attains a global minimum by the extreme value theorem.

Next, we show that in that extended problem for all design parameters, i.e., $b_{k,j}, c_{k,j}, d_{k,j} \in \bar{\mathbb{R}}$, there are finite design parameters $\tilde{b}_{k,j}, \tilde{c}_{k,j}, \tilde{d}_{k,j} \in \mathbb{R}$ achieving the same cost. Consider the action of receiver at stage $k \in \{1, \cdots, n\}$, i.e., $\mathbb{E}\{x_k | y_1, \cdots, y_j, \cdots, y_k\}$, where $y_j$ is given by (11) for each $j = 1, \cdots, k$. If more than one of the design parameters from the set $\{b_{j,1}, ..., b_{j,j}, c_{j,1}, ..., c_{j,j}, d_{j,1}, ..., d_{j,j}\}$, or just one of $c_{j,i}$ or $d_{j,i}$, where $i \in \{1, \cdots, j\}$, is not finite, then the action can also be written as

$$\mathbb{E}\{x_k | y_1, ..., y_j, ..., y_k\} = \mathbb{E}\{x_k | y_1, ..., y_{j-1}, y_{j+1}, ..., y_k\},$$

which can also be obtained by setting $b_{j,i} = c_{j,i} = d_{j,i} = 0$, $i = 1, \cdots, j$. If just one of $b_{j,i}$, $i \in \{1, \cdots, j\}$, is not finite,

say $b_{j,i}$, then
$$\mathbb{E}\{x_k|y_1,...,y_j,...,y_k\} = \mathbb{E}\{x_k|y_1,...,y_{j-1},x_i,y_{j+1},...,y_k\},$$
which can also be obtained by setting $b_{j,i} = 1$ and the other design parameters to zero, i.e., $b_{j,t} = 0 \; \forall t \in \{1,\cdots,j\} \setminus \{i\}$ and $c_{j,i} = d_{j,i} = 0$ for $i = 1,\cdots,j$. Since we choose arbitrary $k \in \{1,\cdots,n\}$ and $j \in \{1,\cdots,k\}$, for the optimum design parameters, i.e., $b_{k,j}^*, c_{k,j}^*, d_{k,j}^* \in \bar{\mathbb{R}}$, there are finite design parameters $\tilde{b}_{k,j}^*, \tilde{c}_{k,j}^*, \tilde{d}_{k,j}^* \in \mathbb{R}$ achieving the same cost. Hence, for the design parameters chosen from the real line, the objective function (14) attains the global minimum. □

Let us first recall the result in the static setting. By (2) and (8), the single-stage, i.e., $n = 1$, Stackelberg equilibrium problem is given by
$$\min_{b_{1,1},c_{1,1},d_{1,1}\in\mathbb{R}} \mathbb{E}\{(x_1+\theta_1 - \mathbb{E}\{x_1|b_{1,1}x_1+c_{1,1}\theta_1+d_{1,1}z_1\})^2\}. \quad (15)$$
Let $x_1$ and $\theta_1$ be correlated, i.e., $\rho := \mathbb{E}\{x_1\theta_1\} \neq 0$. Then, the following lemma presents the Stackelberg solution of the single-state problem.

**Lemma 1.** [16] *For the single-stage Stackelberg equilibrium problem* (15), *essentially unique optimal design parameters of sender*, $(b_{1,1}^*, c_{1,1}^*, d_{1,1}^*)$, *are*
$$b_{1,1}^* = 1, \; c_{1,1}^* = \frac{-\sigma_x^2 + \sigma_x\sqrt{\sigma_x^2+4(\sigma_\theta^2+\rho)}}{2(\sigma_\theta^2+\rho)}, \; d_{1,1}^* = 0. \quad (16)$$

Next, let us consider the two-stage equilibrium, which can also be written as
$$\min_{\substack{b_{1,1},c_{1,1},d_{1,1},\\b_{2,1},c_{2,1},d_{2,1},\\b_{2,2},c_{2,2},d_{2,2}\in\mathbb{R}}} \mathbb{E}\{(x_1+\theta_1-\mathbb{E}\{x_1|b_{1,1}x_1+c_{1,1}\theta_1+d_{1,1}z_1\})^2\}$$
$$+ \mathbb{E}\bigg\{\bigg(x_2+\theta_2-\mathbb{E}\bigg\{x_2\bigg|b_{1,1}x_1+c_{1,1}\theta_1+d_{1,1}z_1,$$
$$\sum_{j=1}^2 b_{2,j}x_j+c_{2,j}\theta_j+d_{2,j}z_j\bigg\}\bigg)^2\bigg\}. \quad (17)$$

The optimization problem (17) has a finite number of minimization arguments, i.e., 9 arguments for the 2-stage equilibrium problem, and can be solved by standard optimization approaches. However, this is computationally involved; hence for tractability reasons, here we employ a different approach.

The following lemma addresses whether the policies of sender can be memoryless or not in the multi-stage Stackelberg equilibrium problem (14), and we say that a policy $\eta_k$, $k \in \{1,\cdots,n\}$, is memoryless if $\eta_k$ depends solely on $x_k$, $\theta_k$, and $z_k$.

**Lemma 2.** *In the multi-stage Stackelberg equilibrium* (14), *i.e.*, $n > 1$, *the optimal linear policies of sender can be constructed as memoryless.*

*Proof.* By (8), the Stackelberg equilibrium problem (14) can also be written in terms of the design parameters as
$$\min_{\substack{b_{j,i},c_{j,i},d_{j,i}\in\mathbb{R}\\j=1,\cdots,n,i=1,\cdots,j}} \sum_{j=1}^n \mathbb{E}\{(x_j+\theta_j-\mathbb{E}\{x_j|y_1,\cdots,y_j\})^2\}, \quad (18)$$
where $y_j$ depends on the design parameters as in (11). Then, by the Principle of Optimality [21], for the given optimal actions $y_1^*,\cdots,y_{n-1}^*$, the optimal design parameters $b_{n,i}^*, c_{n,i}^*, d_{n,i}^*$, $i = 1,\cdots,n$, with respect to (18) minimize the following sub-problem:
$$\min_{\substack{b_{n,i},c_{n,i},d_{n,i}\in\mathbb{R}\\i=1,\cdots,n}} \mathbb{E}\{(x_n+\theta_n-\mathbb{E}\{x_n|y_1^*,\cdots,y_{n-1}^*,y_n\})^2\}. \quad (19)$$
Let $\hat{x}_n := \mathbb{E}\{x_n|y_1^*,\cdots,y_{n-1}^*\}$ such that (19) can be written as
$$\min_{\substack{b_{n,i},c_{n,i},d_{n,i}\in\mathbb{R}\\i=1,\cdots,n}} \mathbb{E}\{(x_n-\hat{x}_n+\theta_n-\mathbb{E}\{x_n-\hat{x}_n|y_n-\mathbb{E}\{y_n|\hat{x}_n\}\})^2\}$$
$$= \min_{\substack{b_{n,i},c_{n,i},d_{n,i}\in\mathbb{R}\\i=1,\cdots,n}} \mathbb{E}\{(\tilde{x}_n+\tilde{\theta}_n-\mathbb{E}\{\tilde{x}_n|\tilde{y}_n\})^2\} + \mathbb{E}\{(\mathbb{E}\{\theta_n|\hat{x}_n\})^2\}, \quad (20)$$
where we let $\tilde{x}_n := x_n - \hat{x}_n$, $\tilde{\theta}_n := \theta_n - \mathbb{E}\{\theta_n|\hat{x}_n\}$, and $\tilde{y}_n := y_n - \mathbb{E}\{y_n|\hat{x}_n\}$, and in (20), we use the orthogonality of $x_n - \hat{x}_n$, $\theta - \mathbb{E}\{\theta_n|\hat{x}_n\}$, and $y_n - \mathbb{E}\{y_n|\hat{x}_n\}$ with $\hat{x}_n$. Note that $\tilde{y}_n$ is given by
$$\tilde{y}_n = \sum_{i=1}^n b_{n,i}\tilde{x}_i + c_{n,i}\tilde{\theta}_i + d_{n,i}\tilde{z}_i$$
$$= b_{n,n}\tilde{x}_n + c_{n,n}\tilde{\theta}_n + d_{n,n}\tilde{z}_n + \underbrace{\sum_{i=1}^{n-1} b_{n,i}\tilde{x}_i + c_{n,i}\tilde{\theta}_i + d_{n,i}\tilde{z}_i}_{=:\omega_n}, \quad (21)$$
where $\tilde{x}_i = x_i - \mathbb{E}\{x_i|\hat{x}_n\}$, $\tilde{\theta}_i = \theta_i - \mathbb{E}\{\theta_i|\hat{x}_n\}$, and $\tilde{z}_i = z_i - \mathbb{E}\{z_i|\hat{x}_n\}$ for $i = 1,\cdots,n-1$, and $\tilde{z}_n = z_n - \mathbb{E}\{z_n|\hat{x}_n\} = z_n$. Then, by adding and subtracting multiples of $\tilde{x}_n$ and $\tilde{\theta}_n$, we obtain
$$\tilde{y}_n = (b_{n,n}+\lambda_x)\tilde{x}_n + (c_{n,n}+\lambda_\theta)\tilde{\theta}_n + d_{n,n}z_n + \underbrace{\omega_n - \lambda_x\tilde{x}_n - \lambda_\theta\tilde{\theta}_n}_{=:\tilde{\omega}_n}.$$

Our aim is to choose $\lambda_x, \lambda_\theta \in \mathbb{R}$ such that $\tilde{\omega}_n$ is pairwise independent of both $\tilde{x}_n$ and $\tilde{\theta}_n$, and $\tilde{\omega}_n$ is independent of $z_n$ by definition. Since all parameters are jointly Gaussian, uncorrelatedness, i.e., $\mathbb{E}\{\tilde{x}_n\tilde{\omega}_n\} = \mathbb{E}\{\tilde{\theta}_n\tilde{\omega}_n\} = 0$, implies pairwise independence. This leads to
$$\begin{bmatrix} \mathbb{E}\{\tilde{x}_n^2\} & \mathbb{E}\{\tilde{x}_n\tilde{\theta}_n\} \\ \mathbb{E}\{\tilde{x}_n\tilde{\theta}_n\} & \mathbb{E}\{\tilde{\theta}_n^2\} \end{bmatrix} \begin{bmatrix} \lambda_x \\ \lambda_\theta \end{bmatrix} = \begin{bmatrix} \mathbb{E}\{\omega_n\tilde{x}_n\} \\ \mathbb{E}\{\omega_n\tilde{\theta}_n\} \end{bmatrix}, \quad (22)$$
where the matrix on the left hand side of (22) is invertible unless there exists a $\tau > 0$ such that the random variables $\tilde{x}_n$ and $\tilde{\theta}_n$ satisfy $\tilde{x}_n = \tau\tilde{\theta}_n$, which is not possible due to the independent noises $w_n$ and $v_n$. Hence, there exist $\lambda_x, \lambda_\theta \in \mathbb{R}$ such that $\tilde{\omega}_n \perp\!\!\!\perp \tilde{x}_n$, $\tilde{\omega}_n \perp\!\!\!\perp \tilde{\theta}_n$, and $\tilde{\omega}_n \perp\!\!\!\perp z_n$, and we obtain[4]
$$\tilde{y}_n = \underbrace{(b_n+\lambda_x)}_{=\tilde{b}_n}\tilde{x}_n + \underbrace{(c_n+\lambda_\theta)}_{=\tilde{c}_n}\tilde{\theta}_n + \underbrace{d_n z_n + \tilde{\omega}_n}_{=\tilde{d}_{n,n}v_n}, \quad (23)$$
where $\tilde{b}_n, \tilde{c}_n, \tilde{d}_n \in \mathbb{R}$ and $v_n$ is a standard normal random variable that is independent of the other parameters. Since the last term of the objective function in (20) is independent

---
[4] For notational simplicity, we let $b_j = b_{j,j}$, $c_j = c_{j,j}$, and $d_j = d_{j,j}$.

of the minimization arguments, by (23) we can write the minimization problem (20) as

$$\min_{b_n,c_n,d_n \in \mathbb{R}} \mathbb{E}\{(\tilde{x}_n + \tilde{\theta}_n - \mathbb{E}\{\tilde{x}_n|b_n\tilde{x}_n + c_n\tilde{\theta}_n + d_nv_n\})^2\}. \quad (24)$$

This shows that the last stage action of the sender can be chosen as a memoryless policy.

Next, consider the stage-$k$ action of the sender and assume that the policies $\eta_{k+1},\cdots,\eta_n$ are memoryless. Then, for the given optimal actions $y_1^*,\cdots,y_{k-1}^*$, the policies $\eta_k,\cdots,\eta_n$ or the corresponding design parameters, minimize the following sub-problem:

$$\min_{\substack{b_{j,i},c_{j,i},d_{j,i} \in \mathbb{R} \\ j=k,\cdots,n, i=1,\cdots,j}} \sum_{j=k}^n \mathbb{E}\{(x_j + \theta_j - \mathbb{E}\{x_j|y_1^*,\cdots,y_{k-1}^*,\cdots,y_j\})^2\}. \quad (25)$$

Since $\mathbb{E}\{x_j|\mathbb{E}\{x_i|y_1^*,\cdots,y_{k-1}^*\}\} = \mathbb{E}\{x_j|y_1^*,\cdots,y_{k-1}^*\}$ and $\mathbb{E}\{\theta_j|\mathbb{E}\{x_i|y_1^*,\cdots,y_{k-1}^*\}\} = \mathbb{E}\{\theta_j|\mathbb{E}\{x_j|y_1^*,\cdots,y_{k-1}^*\}\}$, $i \geq j \geq k$, by following identical steps with (19)-(23), we can write (25) as

$$\min_{\substack{b_{j,i},c_{j,i},d_{j,i} \in \mathbb{R} \\ j=k,\cdots,n, i=1,\cdots,j}} \sum_{j=k}^n \mathbb{E}\left\{\left(\bar{x}_j + \bar{\theta}_j - \mathbb{E}\{\bar{x}_j|\bar{y}_k,\cdots,\bar{y}_j\}\right)^2\right\}, \quad (26)$$

where for $j = k,\cdots,n$, $\bar{x}_j := x_j - \mathbb{E}\{x_j|y_1^*,\cdots,y_{k-1}^*\}$, $\bar{\theta}_j := \theta_j - \mathbb{E}\{\theta_j|\mathbb{E}\{x_j|y_1^*,\cdots,y_{k-1}^*\}\}$, $\bar{y}_j := b_j\bar{x}_j + c_j\bar{\theta}_j + d_jv_j$, and $v_j$ is a standard normal random variable that is independent of the other parameters. Hence, (26) implies that the optimal policy $\eta_k^* \in \Omega_k$ can be chosen as memoryless. By induction, we conclude that in the multi-stage Stackelberg equilibrium, all the policies of sender can be chosen as memoryless. □

**Remark 1.** *We point out that (24) and Lemma 1 yield that in the multi-stage Stackelberg equilibrium (14), the optimal policy of the sender at the last stage, i.e., $\eta_n^* \in \Omega_n$, is a pure strategy such that the policy includes no noise term, i.e., $d_{n,n}^* = 0$. Furthermore, the corresponding essentially unique design parameters are given by $b_{n,n}^* = 1$ and $0 < c_{n,n}^* < 1$.*

The following theorem addresses whether all the optimal policies of the sender are pure strategies or not.

**Theorem 2.** *The multistage problem, $n > 1$, admits pure strategy Stackelberg equilibrium in the set of linear strategies if, and only if, $a_x = a_\theta$.*

*Proof.* Lemma 2 shows that the optimal linear policies of the sender can be constructed as memoryless. Hence, we consider memoryless policies and we specifically focus on the policy at stage-$(n-1)$. Note that by the Principle of Optimality for given $\eta_1^*,\cdots,\eta_{n-2}^*$, the last two policies, i.e., $\eta_{n-1}$ and $\eta_n$, minimize the following sub-problem:

$$\min_{\substack{\eta_k \in \Omega_k, \\ k=n-1,n}} \sum_{k=n-1}^n \mathbb{E}\left\{\left(x_k + \theta_k - \mathbb{E}\{x_k|y_1 = \eta_1^*,\cdots,\right.\right.$$
$$\left.\left. y_{n-2} = \eta_{n-2}^*,\cdots,y_k = \eta_k\}\right)^2\right\}$$

Then, following analogous steps with (19)-(23), we obtain the following optimization problem:

$$\min_{\substack{b_k,c_k,d_k \in \mathbb{R}, \\ k=n-1,n}} E\{(\tilde{x}_{n-1} + \tilde{\theta}_{n-1} - \mathbb{E}\{\tilde{x}_{n-1}|\tilde{y}_{n-1}\})^2\}$$
$$+ E\{(\tilde{x}_n + \tilde{\theta}_n - \mathbb{E}\{\tilde{x}_n|\tilde{y}_{n-1},\tilde{y}_n\})^2\}, \quad (27)$$

where for $k \in \{n-1, n\}$, we redefine $\tilde{x}_k := x_k - \mathbb{E}\{x_k|y_1^*,\cdots,y_{n-2}^*\}$, $\tilde{\theta}_k := \theta_k - \mathbb{E}\{\theta_k|\mathbb{E}\{x_k|y_1^*,\cdots,y_{n-2}^*\}\}$, $\tilde{y}_k := b_k\tilde{x}_k + c_k\tilde{\theta}_k + d_kv_k$, and $d_n^* = 0$ as shown in Remark 1.

Note that the second term in (27) can be written as

$$\mathbb{E}\{(\tilde{x}_n + \tilde{\theta}_n - \mathbb{E}\{\tilde{x}_n|\tilde{y}_{n-1},\tilde{y}_n\})^2\}$$
$$= \mathbb{E}\{(\tilde{x}_n + \tilde{\theta}_n - \mathbb{E}\{\tilde{x}_n|\tilde{y}_n - \mathbb{E}\{\tilde{y}_{n-1}|\tilde{y}_n\}\})^2\}$$
$$= \mathbb{E}\{(\tilde{x}_n + \breve{\theta}_n - \mathbb{E}\{\tilde{x}_n|\tilde{y}_{n-1} - \mathbb{E}\{\tilde{y}_{n-1}|\tilde{y}_n\}\})^2\}$$
$$+ \mathbb{E}\{(\mathbb{E}\{\tilde{\theta}_n|\tilde{y}_n\})^2\},$$

where $\breve{x}_n := \tilde{x}_n - \mathbb{E}\{\tilde{x}_n|\tilde{y}_n\}$ and $\breve{\theta}_n := \tilde{\theta}_n - \mathbb{E}\{\tilde{\theta}_n|\tilde{y}_n\}$. We also define $\breve{x}_{n-1} := \tilde{x}_{n-1} - \mathbb{E}\{\tilde{x}_{n-1}|\tilde{y}_n\}$ and $\breve{\theta}_{n-1} := \tilde{\theta}_{n-1} - \mathbb{E}\{\tilde{\theta}_{n-1}|\tilde{y}_n\}$ such that $\breve{y}_{n-1} := \tilde{y}_{n-1} - \mathbb{E}\{\tilde{y}_{n-1}|\tilde{y}_n\} = b_{n-1}\breve{x}_{n-1} + c_{n-1}\breve{\theta}_{n-1} + d_{n-1}v_{n-1}$. Then, we have the following minimization problem:

$$\min_{\substack{b_k,c_k,d_k \in \mathbb{R}, \\ k=n-1,n}} \mathbb{E}\{(\tilde{x}_{n-1} + \tilde{\theta}_{n-1} - \mathbb{E}\{\tilde{x}_{n-1}|\tilde{y}_{n-1}\})^2\}$$
$$+ \mathbb{E}\{(\breve{x}_n + \breve{\theta}_n - \mathbb{E}\{\tilde{x}_n|\breve{y}_{n-1}\})^2\} + \mathbb{E}\{(\mathbb{E}\{\tilde{\theta}_n|\tilde{y}_n\})^2\}. \quad (28)$$

We point out that by the orthogonality of the estimated variable and the estimation error, i.e., $(\tilde{x}_{n-1} - \mathbb{E}\{\tilde{x}_{n-1}|\tilde{y}_{n-1}\}) \perp \mathbb{E}\{\tilde{x}_{n-1}|\tilde{y}_{n-1}\}$, we obtain

$$\mathbb{E}\{(\tilde{x}_{n-1} + \tilde{\theta}_{n-1} - \mathbb{E}\{\tilde{x}_{n-1}|\tilde{y}_{n-1}\})^2\} = \mathbb{E}\{(\tilde{x}_{n-1} + \tilde{\theta}_{n-1})^2\}$$
$$- \mathbb{E}\{(2\tilde{\theta}_{n-1} + \tilde{x}_{n-1})\mathbb{E}\{\tilde{x}_{n-1}|\tilde{y}_{n-1}\}\}$$
$$+ \mathbb{E}\{(\tilde{x}_{n-1} - \mathbb{E}\{\tilde{x}_{n-1}|\tilde{y}_{n-1}\})\mathbb{E}\{\tilde{x}_{n-1}|\tilde{y}_{n-1}\}\}$$
$$= \mathbb{E}\{(\tilde{x}_{n-1} + \tilde{\theta}_{n-1})^2\} - \mathbb{E}\{(2\tilde{\theta}_{n-1} + \tilde{x}_{n-1})\mathbb{E}\{\tilde{x}_{n-1}|\tilde{y}_{n-1}\}\}$$
(29)

and correspondingly

$$\mathbb{E}\{(\breve{x}_n + \breve{\theta}_n - \mathbb{E}\{\breve{x}_n|\breve{y}_{n-1}\})^2\} = \mathbb{E}\{(\breve{x}_n + \breve{\theta}_n)^2\}$$
$$- \mathbb{E}\{(2\breve{\theta}_n + \breve{x}_n)\mathbb{E}\{\breve{x}_n|\breve{y}_{n-1}\}\}. \quad (30)$$

By (29) and (30), the optimization problem (28) can be written as

$$\min_{\substack{b_k,c_k,d_k \in \mathbb{R}, \\ k=n-1,n}} \mathbb{E}\{(\tilde{x}_{n-1} + \tilde{\theta}_{n-1})^2\} + \mathbb{E}\{(\breve{x}_n + \breve{\theta}_n)^2\}$$
$$- \mathbb{E}\{(2\tilde{\theta}_{n-1} + \tilde{x}_{n-1})\mathbb{E}\{\tilde{x}_{n-1}|\tilde{y}_{n-1}\}\}$$
$$- \mathbb{E}\{(2\breve{\theta}_n + \breve{x}_n)\mathbb{E}\{\breve{x}_n|\breve{y}_{n-1}\}\} + \mathbb{E}\{(\mathbb{E}\{\tilde{\theta}_n|\tilde{y}_n\})^2\}. \quad (31)$$

Eventually, after some algebra for (31), we have the optimization problem in (32), where for $k \in \{n-1,n\}$, we define

$$\varphi_k = \mathbb{E}\{\tilde{x}_k\tilde{y}_k\} = b_k\mathbb{E}\{\tilde{x}_k^2\} + c_k\mathbb{E}\{\tilde{x}_k\tilde{\theta}_k\}$$
$$\phi_k = \mathbb{E}\{\tilde{\theta}_k\tilde{y}_k\} = b_k\mathbb{E}\{\tilde{\theta}_k\tilde{x}_k\} + c_k\mathbb{E}\{\tilde{\theta}_k^2\}, \quad (33)$$

which are independent of $d_k$, and

$$\Delta := \frac{a_x\phi_n\varphi_{n-1} - a_\theta\phi_{n-1}\varphi_n}{b_n\varphi_n + c_n\phi_n}. \quad (34)$$

$$\max_{\substack{b_k,c_k,d_k\in\mathbb{R}\\k=n-1,n}} \frac{\varphi_{n-1}(2\phi_{n-1}+\varphi_{n-1})}{b_{n-1}\varphi_{n-1}+c_{n-1}\phi_{n-1}+d_{n-1}^2} + \frac{\varphi_n(2\phi_n+\varphi_n)}{b_n\varphi_n+c_n\phi_n} + \frac{c_n(c_n-2b_n)\Delta^2}{b_{n-1}\varphi_{n-1}+c_{n-1}\phi_{n-1}-\frac{(a_x b_n \varphi_{n-1}+a_\theta c_n \phi_{n-1})^2}{b_n\varphi_n+c_n\phi_n}+d_{n-1}^2} \quad (32)$$

Proposition 1 says that (32) attains globally optimal solutions, and let $b_k^*, c_k^*$, and $d_k^*$ for $k \in \{n-1, n\}$ be the globally optimal solutions of (32). Note that $d_{n-1}^2 \geq 0$ stands only in the first and the last terms in (32), and the quadratic terms,

$$b_{n-1}\varphi_{n-1} + c_{n-1}\phi_{n-1} - \frac{(a_x b_n \varphi_{n-1} + a_\theta c_n \phi_{n-1})^2}{b_n\varphi_n + c_n\phi_n}$$
$$= \mathbb{E}\{(b_{n-1}\check{x}_{n-1} + c_{n-1}\check{\theta}_{n-1})^2\}, \quad (35)$$
$$b_{n-1}\varphi_{n-1} + c_{n-1}\phi_{n-1} = \mathbb{E}\{(b_{n-1}\tilde{x}_{n-1} + c_{n-1}\tilde{\theta}_{n-1})^2\}, \quad (36)$$

are non-negative. Assume that $d_{n-1}^* = 0$. Due to the quadratic terms (35) and (36), as seen in (32), $d_{n-1}^* = 0$ implies that the first and the last term in (32) are non-negative. However, by (24), Lemma 1 and Remark 1, the essentially unique optimal design parameters in the last stage action are given by $b_n^* = 1$, $0 < c_n^* < 1$, and $d_n^* = 0$, which yields that the last term in (32) is non-positive and it is zero if $\Delta = 0$. Hence, if $d_{n-1}^* = 0$, the other optimal design parameters satisfy $\Delta = 0$. Then, for given $\Delta = 0$ and $d_{n-1}^* = 0$, (32) can be written as

$$\max_{b_k,c_k\in\mathbb{R},k\in\{n-1,n\}} \sum_{k=n-1}^{n} \frac{\varphi_k(2\varphi_k+\phi_k)}{b_k\varphi_k+c_k\phi_k},$$

where the objective function is separable with respect to optimization arguments such that

$$\max_{b_k,c_k\in\mathbb{R}} \frac{\varphi_k(2\varphi_k+\phi_k)}{b_k\varphi_k+c_k\phi_k}, \quad (37)$$

for $k \in \{n-1, n\}$. The optimal action of the sender at stage-$(n-1)$ has no impact on the optimal action at stage-$n$ even for the finite horizon objective. Given that the objective function is separable for the last $(m-1)$-stages, in a similar way, it can be shown that the action at stage-$m$ has no impact on the future stages. Hence, by induction, we conclude that the objective function (14) is separable with respect to the optimization arguments as in (37) for $k \in \{1, \cdots, n\}$, and $d_k^* = 0$ for all $k \in \{1, \cdots, n\}$. However, this would imply that the first two stages can also be separated, which yields that

$$a_x b_2 c_1 = a_\theta b_1 c_2 \quad (38)$$

by (33) and (34). Correspondingly, because of the separability of the objective function, (14) can be separated as

$$\min_{\eta_k \in \Omega_k} \mathbb{E}\{(x_k + \theta_k - \mathbb{E}\{x_k|y_k\})^2\}$$

for $k = 1, \cdots, n$, which leads to $\eta_1^* = \eta_2^* = \cdots = \eta_n^*$ due to the stationarity of the state $x_k$ and the bias $\theta_k$. Then, (38) yields that $a_x = a_\theta$, which is a contradiction since we consider any $a_x, a_\theta \in (-1, 1)$ a priori. Hence, $d_{n-1}^* = 0$ if and only if $a_x = a_\theta$, and if $a_x = a_\theta$, then $d_k^* = 0$ for all $k = 1, \cdots, n$. $\square$

## V. CONCLUSIONS

In this paper, we have analyzed the structure of linear equilibrium strategies for dynamic Gaussian signaling games. Our main result is that unless the state and bias processes evolve in an identical fashion, a pure (linear) strategy does not achieve the Stackelberg equilibrium for a dynamic game, while the static version of the game admits pure (linear) strategies as the essentially unique solution. This difference in the optimal strategies is due to the trade-off between costs at different stages of the game; such trade-offs were also encountered in several other asymmetric information games. We have limited the space of strategies to linear; the analysis in unrestricted policy spaces is left as part of future work.

## REFERENCES

[1] V. Crawford and J. Sobel, "Strategic information transmission," *Econometrica: Journal of the Econometric Society*, pp. 1431–1451, 1982.
[2] M. Spence, "Job market signaling," *The Quarterly Journal of Economics*, pp. 355–374, 1973.
[3] M. Gentzkow and E. Kamenica, "Bayesian persuasion," *American Economic Review*, vol. 101, no. 6, pp. 2590–2615, 2011.
[4] L. Rayo and I. Segal, "Optimal information disclosure," *Journal of Political Economy*, vol. 118, no. 5, pp. 949–987, 2010.
[5] M. Gentzkow and E. Kamenica, "Costly persuasion," *The American Economic Review*, vol. 104, no. 5, pp. 457–462, 2014.
[6] ——, "A Rothschild-Stiglitz approach to Bayesian persuasion," in *American Economic Review Papers & Proceedings*, 2015.
[7] J. Sobel, "Giving and receiving advice," in *Econometric Society 10th World Congress*, 2010.
[8] M. Gentzkow and E. Kamenica, "Competition in persuasion," National Bureau of Economic Research, Tech. Rep., 2011.
[9] D. Bergemann and S. Morris, "Robust predictions in games with incomplete information," *Econometrica*, vol. 81, no. 4, pp. 1251–1308, 2013.
[10] Y. Che, W. Dessein, and N. Kartik, "Pandering to Persuade," *The American Economic Review*, vol. 103, no. 1, pp. 47–79, 2013.
[11] E. Perez-Richet, "Interim bayesian persuasion: First steps," *The American Economic Review*, vol. 104, no. 5, pp. 469–474, 2014.
[12] A. Kolotilin, "Experimental design to persuade," *Games and Economic Behavior*, vol. 90, pp. 215–226, 2015.
[13] F. Farokhi, A. Teixeira, and C. Langbort, "Estimation with strategic sensors," *IEEE Transactions on Automatic Control*, in press.
[14] S. Sarıtaş, S. Yüksel, and S. Gezici, "Quadratic multi-dimensional signaling games and affine equilibria," *IEEE Transactions on Automatic Control*, in press.
[15] E. Akyol, C. Langbort, and T. Başar, "Strategic compression and transmission of information," in *Proceedings of the IEEE Inf. Theory Workshop, Korea*, 2015.
[16] ——, "Information-theoretic approach to strategic communication as a hierarchical game," *Proceedings of the IEEE*, in press.
[17] A. Gupta, A. Nayyar, C. Langbort, and T. Başar, "Common information based Markov perfect equilibria for linear-Gaussian games with asymmetric information," *SIAM Journal on Control and Optimization*, vol. 52, no. 5, pp. 3228–3260, 2014.
[18] W. Tamura, "A theory of multidimensional information disclosure," *Working paper, available at SSRN 1987877*, 2014.
[19] J. Harsanyi, "Games with incomplete information played by Bayesian players, I-III: Part I. The basic model," *Management Science*, vol. 50, no. 12_supplement, pp. 1804–1817, 2004.
[20] T. Başar and G. Olsder, *Dynamic Noncooperative Game Theory*. Society for Industrial Mathematics (SIAM) Series in Classics in Applied Mathematics, 1999.
[21] D. Bertsekas, *Dynamic programming and optimal control*. Athena Scientific Belmont, MA, 1995, vol. 1, no. 2.